# An Autonomous Scientific Knowledge Generation Framework for AI-Driven Scientific Discovery

**Dibakar Datta**

Dibakar Datta Laboratory (DDLab), Department of Mechanical and Industrial Engineering
New jersey Institute of Technology (NJIT), Newark, NJ 07052, USA
Email: ddlab@njit.edu ; Phone: +1 973-596-3647

---

## Abstract

Artificial intelligence (AI) is rapidly transforming scientific discovery. However, its effectiveness remains fundamentally constrained by the availability of high-quality, structured scientific knowledge (data). Although community databases have accelerated data-driven materials research, much of the scientific knowledge required for predictive modeling and inverse materials design remains embedded within unstructured literature, including narrative text, tables, figures, equations, and supplementary materials. Transforming this heterogeneous information into machine-interpretable knowledge remains a major challenge for next-generation AI-driven scientific discovery. Here, we present an *Autonomous Scientific Knowledge Generation Framework* that systematically transforms scientific literature into a *Unified AI-Ready Scientific Knowledge Base*. The framework integrates ontology-guided literature acquisition, hybrid scientific knowledge extraction, semantic harmonization, evidence-based knowledge fusion, and ontology-guided validation within a unified computational architecture. Rather than treating literature retrieval, information extraction, and database construction as independent tasks, the proposed framework progressively converts distributed scientific observations into structured, semantically consistent, provenance-preserving knowledge suitable for downstream AI-driven reasoning. The framework is demonstrated using electro-optic materials as a representative benchmark because of their complex tensor properties, multidimensional operating conditions, and heterogeneous reporting conventions. Autonomous literature acquisition retrieved and validated approximately 1,000 'best' domain-specific publications from multiple scholarly repositories. A representative subset of eight publications was subsequently processed through the complete workflow, generating 29 structured scientific records, which were harmonized into 7 canonical scientific records. The demonstration illustrates the complete transformation from scientific publications to structured knowledge and ultimately to an AI-ready knowledge base while preserving quantitative measurements together with their associated materials, operating conditions, provenance, and scientific context. Beyond literature mining, the proposed framework establishes a general architecture for autonomous scientific knowledge generation. By integrating literature-derived knowledge with computational simulations, experimental measurements, existing scientific repositories, and future observations, the framework provides the information infrastructure required for predictive scientific intelligence, generative scientific intelligence, and closed-loop autonomous scientific discovery. Because the underlying architecture is domain agnostic, it can be readily adapted to batteries, quantum materials, catalysis, semiconductors, polymers, biomaterials, nanomedicine, and other data-intensive scientific disciplines, providing a scalable foundation for next-generation AI-assisted scientific discovery.

## 1. INTRODUCTION

Artificial Intelligence (AI) is transforming the way scientific knowledge is generated, analyzed, and utilized[1, 2]. Advances in machine learning (ML)[3], natural language processing (NLP)[4], deep learning (DL)[5], large language models (LLMs)[6], and generative AI[7] have enabled computational systems to recognize complex patterns, integrate heterogeneous information, perform scientific reasoning, and assist in hypothesis generation, experimental planning, and materials design. Beyond serving as analytical tools, modern AI systems are increasingly becoming active participants in scientific research, enabling a transition from data-driven computation toward AI-assisted scientific discovery[8]. Their impact now spans diverse disciplines, including computer vision[9], molecular and protein design[10], drug discovery[11], and materials science[12], demonstrating that AI can not only analyze existing knowledge but also generate new scientific hypotheses and candidate designs[13, 14]. Among these applications, materials science represents one of the most promising domains for AI-enabled discovery[14]. Traditionally, the development of functional materials has relied on iterative cycles of theoretical modeling, computational simulation, experimental synthesis, characterization, and optimization, often requiring years of sustained effort before technologically useful materials are identified. The immense combinatorial design space arising from chemical composition, crystal structure, microstructure, processing history, and external operating conditions makes exhaustive experimental or computational exploration infeasible. Recent advances in AI are beginning to fundamentally reshape this paradigm by enabling rapid property prediction, large-scale virtual screening, inverse materials design[13], and autonomous experimentation[15, 16]. Combined with high-throughput simulations, robotics, and automated laboratories, these technologies are accelerating the transition from conventional trial-and-error materials development toward intelligent, data-driven scientific discovery.

Despite these advances, the performance of AI models remains fundamentally constrained by the quality, completeness, and consistency of the scientific knowledge (data) available for learning. Regardless of algorithmic sophistication, every AI model ultimately learns only from the training data. Consequently, the principal bottleneck in next-generation AI-enabled materials discovery is no longer the development of increasingly powerful learning algorithms alone, but the construction of comprehensive, reliable, and continuously evolving scientific knowledge capable of representing the complex relationships among materials composition, crystal structure, processing history, external operating conditions, computational methodologies, and functional performance. Fragmented or incomplete knowledge limits not only prediction accuracy but also the ability of AI systems to generalize beyond previously explored materials systems. Community resources such as the Materials Project[17], the Open Quantum Materials Database (OQMD)[18], Automatic-FLOW for Materials Discovery (AFLOW)[19], and the Inorganic Crystal Structure Database (ICSD)[20] have established the data foundation for modern computational materials science by providing standardized crystal structures, formation energies, electronic properties, elastic constants, phonons, and other experimentally measured or computationally derived descriptors. These databases have dramatically accelerated high-throughput screening and machine-learning-based materials discovery. However, they capture only part of the scientific knowledge required for comprehensive AI-driven discovery. Functional materials behavior frequently depends on synthesis protocols[21], processing history, defects, crystallographic orientation[22], phase composition[23], measurement methodology, uncertainty, and external operating conditions[21] such as temperature, pressure, electric field[21], magnetic field, mechanical strain[24], wavelength, and chemical environment. Such contextual information is often inseparable from the reported material properties themselves but remains largely absent from conventional structured databases.

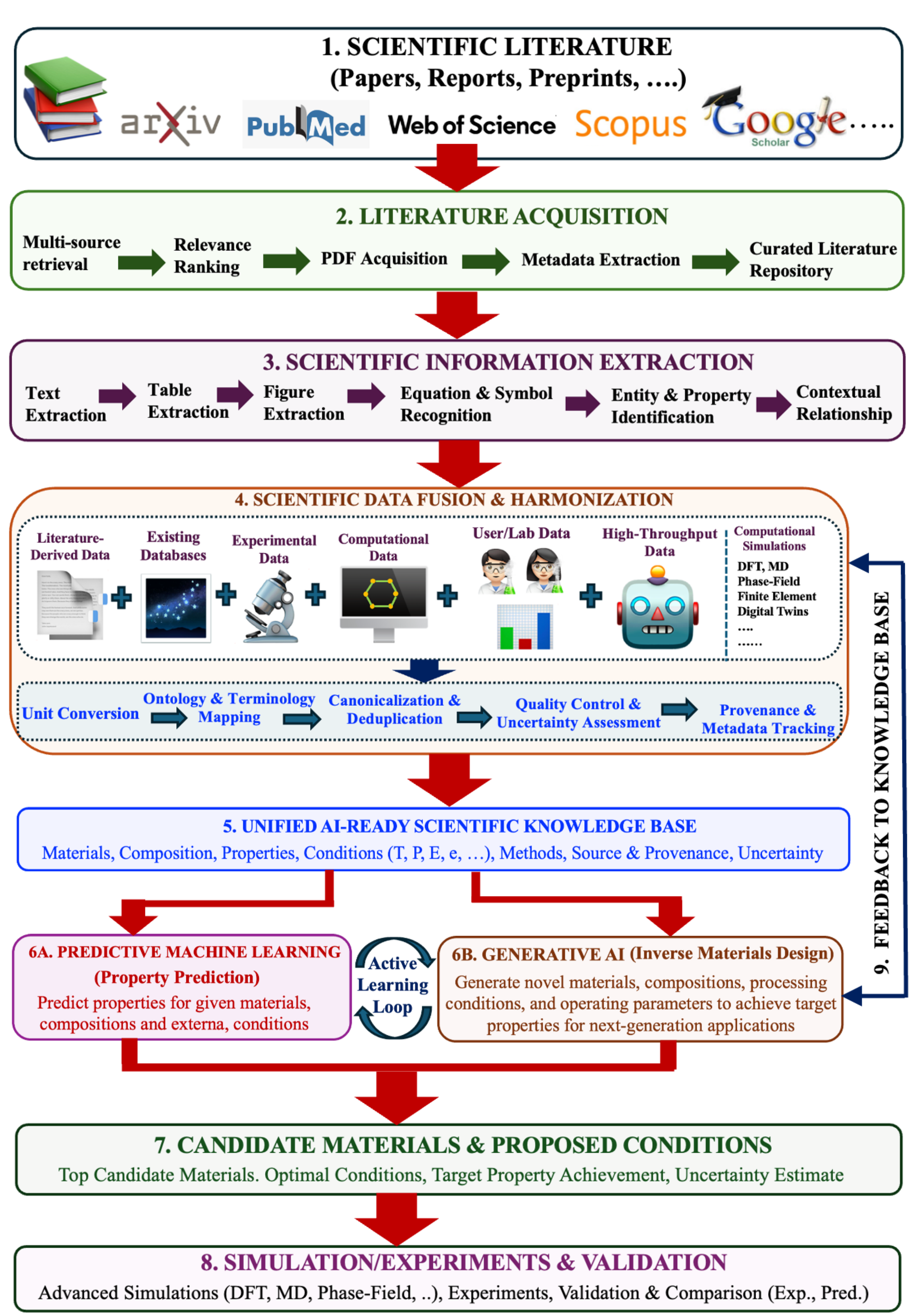

**Figure 1: Autonomous Scientific Data Generation Framework**

The resulting knowledge gap becomes increasingly significant for modern functional materials, whose performance is governed by strongly coupled physical phenomena spanning multiple length and time scales. Whether designing next generation batteries[7, 25], quantum materials[21], electro-optic devices, catalysts, semiconductors, or biomaterials[26], AI models require far more than isolated material descriptors. They require integrated scientific knowledge that preserves the relationships among materials, processing conditions, computational methodologies, experimental observations, and resulting functional properties[27]. Fortunately, much of this knowledge already exists. Over several decades, researchers have generated an enormous body of scientific information through experiments, simulations, and theoretical studies. Collectively, these publications constitute humanity's largest continuously expanding scientific knowledge repository, documenting materials compositions, processing strategies, characterization methods, computational results, functional properties, and operating conditions across virtually every area of materials science. Unlike conventional materials databases[17, 18], the scientific literature was designed for communication among researchers rather than direct use by artificial intelligence. Scientific knowledge is distributed across text, tables, figures, equations, captions, supplementary files, and graphical representations, where quantitative measurements are embedded within experimental methods, theoretical assumptions, scientific interpretation, and domain-specific terminology. Much of the most valuable knowledge is therefore not available as structured data but must be reconstructed from context. The literature is not merely an unstructured collection of documents; it is a human-readable knowledge system whose organization remains largely inaccessible to machine reasoning.

Transforming this literature into AI-ready knowledge requires more than optical character recognition, keyword search, or numerical extraction. Scientific concepts are reported using diverse nomenclature, abbreviations, symbols, tensor conventions, coordinate systems, units, and measurement protocols. Identical materials may appear as chemical formulas, common names, acronyms, or compositional descriptions, while experimental conditions may be reported numerically, qualitatively, or implicitly. A property value becomes scientifically meaningful only when linked to its material composition, structure, synthesis route, measurement method, computational assumptions, operating temperature, pressure, strain, electric field, wavelength, uncertainty, and provenance[21]. Recovering these relationships requires contextual scientific reasoning rather than conventional text mining. Manual curation cannot solve this problem at scale. The scientific literature is expanding too rapidly for researchers to systematically identify relevant publications, extract contextual measurements, verify reported values, reconcile inconsistent terminology, normalize heterogeneous units, and organize the resulting knowledge into reusable databases. Even in a narrow research area, constructing a reliable database can require months or years of expert effort. Extending this process across materials classes or scientific disciplines is effectively impossible. The central bottleneck is therefore not the absence of scientific knowledge, but the absence of autonomous systems capable of continuously converting literature into structured, context-aware, AI-ready knowledge.

Recent advances in natural language processing[4], large language models, and scientific foundation models have improved literature retrieval, entity recognition, document summarization, relation extraction, question answering, and targeted information extraction. These tools are valuable, but most remain document centric. They help researchers find, read, summarize, or query papers, yet they generally do not construct evolving scientific knowledge bases that integrate heterogeneous information across large publication collections while preserving the relationships among materials, processing conditions, methods, assumptions, and functional properties. The next step is therefore a transition from document mining to scientific knowledge engineering. Here, we propose an *Autonomous Scientific Knowledge Generation Framework* that unifies literature acquisition, scientific knowledge extraction, semantic harmonization, data fusion, and knowledge-base construction within a continuously evolving workflow. As illustrated in **Figure 1**, the framework begins by acquiring scientific literature from multiple scholarly repositories, then extracts materials compositions, processing conditions, experimental observations, computational methods, external stimuli, and functional properties. These literature-derived records are subsequently fused with existing databases, simulations, high-throughput calculations, experimental measurements, laboratory-generated data, and other trusted sources. Through unit normalization, ontology mapping, duplicate resolution, provenance tracking, and quality control, these heterogeneous information streams are transformed into a *Unified AI-Ready Scientific Knowledge Base* that preserves both quantitative measurements and the scientific context required for reliable interpretation.

The *Unified AI-Ready Scientific Knowledge Base* provides the foundation for AI-driven scientific reasoning. Predictive machine learning models can utilize this integrated knowledge to estimate materials properties, identify relationships among composition, processing conditions, and performance, and rapidly explore multidimensional design spaces. Building upon these predictive capabilities, generative AI models can perform inverse materials design by proposing new materials together with candidate processing conditions and operating parameters that satisfy predefined functional objectives. Candidate designs can subsequently be validated through first-principles calculations, molecular simulations, high-throughput experiments, or laboratory testing, with the resulting data continuously incorporated back into the knowledge base. This iterative feedback mechanism enables a continuously evolving scientific ecosystem that progressively improves predictive and generative intelligence as new knowledge becomes available. The significance of the proposed framework extends well beyond automated literature mining. By integrating scientific knowledge from literature, simulations, experiments, existing databases, and future observations into a unified knowledge infrastructure, the framework establishes a scalable foundation for predictive machine learning, generative AI, autonomous experimentation, and ultimately AI-driven scientific discovery across diverse scientific disciplines.

To validate the framework, we employ electro-optic materials[28 24, 29, 30] as a representative benchmark rather than as the primary focus of this study. Electro-optic materials[21, 22, 31-33] provide a particularly demanding test case because their behavior depends on complex relationships among crystal symmetry, tensor properties, operating wavelength, temperature, strain, electric field, synthesis history, and measurement methodology. Successfully extracting and harmonizing such multidimensional scientific knowledge demonstrates the ability of the proposed framework to capture rich contextual relationships that are equally important in batteries, catalysis, semiconductors, quantum materials, polymers, biomaterials, nanomedicine[11], and many other scientific domains.

The principal contributions of this work are as follows. First, we introduce an autonomous AI framework that unifies literature acquisition, scientific knowledge extraction, semantic harmonization, data fusion, and knowledge-base construction within a single computational architecture. Second, we introduce the concept of a *Unified AI-Ready Scientific Knowledge Base*, where literature-derived knowledge is continuously integrated with computational, experimental, and existing database resources to create an evolving infrastructure for AI-driven discovery. Third, we develop a modular, domain-agnostic architecture based on configurable ontologies and extraction schemas that can be readily adapted across diverse scientific disciplines. Fourth, we demonstrate the framework using electro-optic materials as a challenging benchmark for autonomous scientific knowledge generation. Finally, we establish a scalable pathway toward closed-loop AI-driven scientific discovery by connecting autonomous knowledge generation with predictive machine learning, generative AI, computational validation, and experimental feedback. Collectively, this work establishes a general methodology for transforming distributed scientific literature into reusable AI-ready scientific knowledge, providing the knowledge infrastructure required for the next generation of autonomous scientific discovery.

## 2. AUTONOMOUS SCIENTIFIC KNOWLEDGE GENERATION FRAMEWORK

The proposed framework establishes a unified computational architecture for autonomous scientific knowledge generation and AI-driven materials discovery. Rather than treating literature retrieval, information extraction, data integration, machine learning, computational modeling, and experimental validation as independent research activities, *the framework integrates these components into a continuously evolving scientific knowledge ecosystem*. As illustrated in **Figure 1**, heterogeneous scientific information is progressively transformed into a *Unified AI-Ready Scientific Knowledge Base*, providing the foundation for predictive scientific intelligence, generative AI, and ultimately autonomous scientific discovery. The workflow begins with autonomous literature acquisition, where ontology-guided retrieval strategies identify scientifically relevant publications from multiple scholarly repositories. Retrieved publications undergo automated document acquisition, metadata validation, and semantic relevance assessment to construct a curated scientific literature repository. The resulting documents are subsequently processed by the *Scientific Knowledge Extraction Module*, which interprets text, tables, figures, equations, and supplementary materials to recover not only scientific entities but also the relationships among materials composition, crystal structure, synthesis history, computational methodologies, operating conditions, external stimuli, and measured functional properties. Unlike conventional information extraction approaches, the framework preserves the scientific context associated with every observation.

The extracted knowledge is then transformed through the *Scientific Data Fusion and Harmonization Module*. Rather than relying exclusively on literature-derived information, the framework integrates scientific knowledge from publications, existing databases, computational simulations, high-throughput calculations, experimental measurements, laboratory-generated data, and future observations. These heterogeneous sources are reconciled through ontology mapping, semantic normalization, unit harmonization, duplicate resolution, provenance tracking, and quality assurance to construct a continuously evolving *Unified AI-Ready Scientific Knowledge Base*. This knowledge base preserves not only quantitative measurements but also the experimental, computational, and contextual information required for reliable scientific interpretation. The unified knowledge base provides the information infrastructure for both predictive and generative scientific intelligence. Predictive machine learning models learn relationships among composition, structure, processing history, operating conditions, and functional performance to rapidly estimate materials properties across large design spaces. Building upon this capability, generative AI performs inverse scientific design by proposing new materials together with candidate processing conditions and operating parameters that satisfy predefined functional objectives. Candidate materials can then be evaluated using first-principles calculations, molecular simulations, high-throughput experiments, or laboratory validation.

Unlike conventional sequential workflows, the proposed framework operates as a continuously learning ecosystem. Newly generated computational and experimental observations are automatically incorporated into the knowledge base, enabling periodic refinement of both predictive and generative AI models. This iterative feedback mechanism transforms the framework from a static literature-mining pipeline into a self-improving scientific knowledge generation system capable of continuously expanding and refining its scientific understanding. Although this work demonstrated using electro-optic materials, the framework is inherently domain agnostic. By modifying the underlying scientific ontology, retrieval strategy, and extraction schema, the same architecture can be readily adapted to batteries, catalysis, semiconductors, quantum materials, polymers, biomaterials, nanomedicine, and numerous other scientific disciplines. Consequently, the proposed framework provides a general foundation for autonomous scientific knowledge generation and closed-loop AI-driven scientific discovery.

# 3. LITERATURE ACQUISITION MODULE

## 3.1 Framework Overview

The *Literature Acquisition Module* constitutes the first component of the proposed *Autonomous Scientific Knowledge Generation Framework*. Its objective is to autonomously construct a high-quality, domain-specific scientific literature repository that serves as the foundation for subsequent knowledge extraction, data fusion, and AI-driven scientific discovery[7]. Instead of relying on manual identification, screening, and downloading of publications, the framework continuously discovers, evaluates, acquires, and organizes scientifically relevant literature through an end-to-end automated workflow. As illustrated in **Figure 2**, the module integrates ontology-guided query generation, multi-source literature retrieval, AI-based relevance ranking, automated PDF acquisition, metadata validation, and repository construction within a unified workflow. These components operate sequentially as a progressive filtering process, transforming a large and heterogeneous collection of publications into a curated repository of scientifically relevant, validated full-text articles.

Quality control is incorporated throughout the acquisition process rather than applied after document collection. Candidate publications undergo semantic relevance assessment before download, while acquired documents are automatically validated, deduplicated, and linked to standardized metadata and provenance information. This progressive refinement substantially reduces irrelevant or low-quality publications while maintaining broad coverage of the target scientific domain. The final output is not simply a collection of downloaded PDF files, but a curated scientific literature repository containing validated publications together with standardized metadata, document identifiers, and provenance records. This repository provides the direct input to the *Scientific Knowledge Extraction Module*, where the acquired literature is transformed into structured, machine-interpretable scientific knowledge.

## 3.2 Ontology-Driven Query Generation

The effectiveness of autonomous literature acquisition depends on generating retrieval queries that accurately represent the target scientific domain. Conventional keyword-based searches often suffer from incomplete retrieval because scientific concepts are expressed using diverse nomenclature, abbreviations, tensor conventions, and domain-specific terminology. Consequently, relying on a small set of manually selected keywords may overlook a substantial portion of the relevant literature. To address this challenge, the proposed framework employs an ontology-driven query generation strategy that translates a scientific research objective into a diverse set of semantically related search queries. Rather than relying solely on isolated keywords, the framework constructs queries from a domain ontology describing the relationships among materials systems, functional properties, experimental techniques, computational methods, operating conditions, and application domains. This ontology enables the retrieval process to capture conceptual rather than purely lexical relationships.

For the electro-optic materials benchmark[21], the ontology includes candidate materials, electro-optic properties, tensor components, crystallographic terminology, characterization techniques, computational methods, synthesis approaches, device applications, and external operating conditions such as temperature, electric field, wavelength, and mechanical strain. These concepts are automatically expanded using scientific synonyms, alternative nomenclature, and common abbreviations to maximize retrieval coverage. Instead of generating a single search expression, the framework produces multiple complementary query categories targeting different aspects of the scientific domain, including materials, functional properties, experimental studies, computational investigations, processing conditions, and device applications. This diversified retrieval strategy substantially improves literature recall while maintaining scientific relevance, establishing the semantic foundation for the subsequent stages of literature retrieval, AI-based relevance ranking, and scientific knowledge extraction.

### 3.3 Multi-Source Literature Retrieval

Scientific knowledge is distributed across diverse publication sources, including peer-reviewed journals, preprint servers, conference proceedings, institutional repositories, and domain-specific databases. Because no single repository provides comprehensive coverage, particularly for interdisciplinary research, the proposed framework employs a multi-source literature retrieval strategy to maximize coverage while minimizing retrieval bias. Following ontology-driven query generation, semantically expanded queries are executed independently across multiple scholarly repositories.

The current implementation integrates Crossref[34], arXiv[35], and PubMed[36], while the modular architecture readily supports additional sources such as Web of Science[37], Scopus[38], Semantic Scholar[39], Google Scholar[40], publisher repositories, and domain-specific databases. Candidate publications retrieved from different sources are aggregated into a unified collection together with standardized metadata, including titles, authors, publication year, abstracts, digital object identifiers (DOIs)[41], source repository, and persistent document identifiers.

Rather than aggressively filtering publications during retrieval, the framework intentionally prioritizes high recall to maximize coverage of the target scientific domain. The resulting candidate collection therefore includes publications from multiple scientific communities, publication types, and research perspectives. Subsequent AI-based relevance ranking progressively refines this broad collection into a curated literature repository, combining comprehensive retrieval with intelligent semantic filtering to establish a robust foundation for downstream scientific knowledge generation.

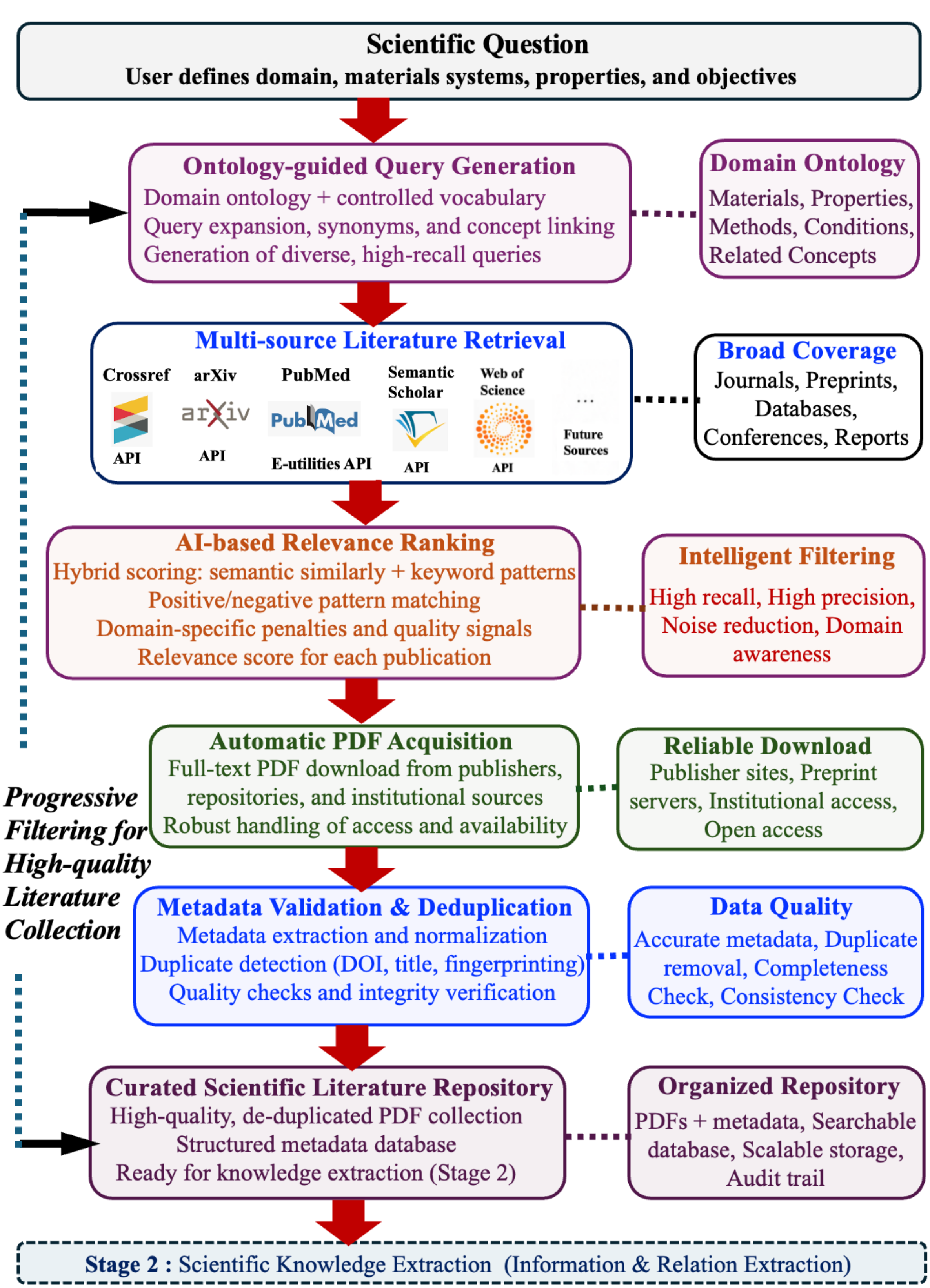

**Figure 2: Literature Acquisition Module**

### 3.4 AI-Based Scientific Relevance Ranking

Following multi-source retrieval, the framework performs AI-based scientific relevance ranking to distinguish publications that are likely to contribute meaningful scientific knowledge from those that are only marginally related to the target domain. Because the retrieval stage intentionally prioritizes high recall, the candidate collection may contain irrelevant or weakly related publications. The relevance-ranking module therefore serves as the primary quality-control mechanism before downstream knowledge extraction. Unlike conventional ranking methods based primarily on keyword frequency or lexical similarity, the proposed framework combines semantic understanding with domain-specific scientific knowledge. Each publication is evaluated using multiple complementary criteria, including semantic similarity, occurrence of domain-specific concepts, publication metadata, and positive and negative semantic patterns that distinguish relevant scientific content from unrelated topics. This hybrid scoring strategy enables the framework to recognize scientifically meaningful publications while reducing false positives arising from ambiguous terminology or overlapping vocabulary.

Rather than assigning binary labels, the framework computes a continuous relevance score that estimates each publication's contribution to the target knowledge base. This allows flexible threshold selection depending on the application, ranging from high-recall exploratory searches to highly curated literature collections for large-scale knowledge generation. By progressively filtering publications according to scientific rather than purely lexical relevance, the ranking module substantially improves the quality of the curated literature repository while preserving the broad coverage required for autonomous scientific knowledge generation.

### 3.5 Automated PDF Acquisition and Repository Construction

Following AI-based relevance assessment, publications satisfying the predefined relevance criteria are automatically acquired through the *PDF Acquisition Module*. Rather than simply downloading documents, this stage constructs a validated, traceable, and reproducible scientific literature repository that serves as the direct input for downstream knowledge extraction. The framework automatically retrieves full-text publications from publisher websites, institutional repositories, open-access archives, and other authorized sources whenever available. During acquisition, document integrity is continuously verified to identify incomplete downloads, corrupted files, inaccessible documents, unsupported formats, and other acquisition errors. Only validated publications are retained for subsequent processing.

For each successfully acquired document, the framework extracts and validates standardized bibliographic metadata, including article title, authors, publication year, journal, digital object identifier (DOI), source repository, document URL, and provenance information. Automated duplicate detection further ensures repository integrity by identifying publications retrieved from multiple sources or repeated across successive acquisition cycles. Consequently, every publication is uniquely represented while preserving complete acquisition provenance and reproducibility. The validated documents and associated metadata are organized into a *Curated Scientific Literature Repository*, where each publication is uniquely identified and linked to standardized metadata and provenance records. This repository provides a reliable and continuously expandable foundation for the subsequent *Scientific Knowledge Extraction Module*, enabling large-scale AI-assisted analysis of scientific literature.

### 3.6 Workflow Orchestration

The individual components of the *Literature Acquisition Module* are coordinated through an automated workflow orchestration engine that enables end-to-end execution of the complete acquisition pipeline. The orchestration layer manages ontology-driven query generation, multi-source literature retrieval, AI-based relevance ranking, automated PDF acquisition, metadata validation, duplicate detection, and repository construction within a single integrated workflow, eliminating the need for manual intervention between successive stages. In addition to coordinating execution, the orchestration engine manages the exchange of standardized metadata, document identifiers, relevance scores, provenance information, and retrieval statistics across all modules. This ensures consistency, traceability, and reproducibility throughout the acquisition process while minimizing processing errors commonly associated with fragmented workflows. Because each component operates as an independent module, the framework can readily incorporate additional retrieval sources, relevance-ranking algorithms, ontologies, or document acquisition strategies without modifying the overall architecture. The modular design also supports distributed execution, incremental repository updates, and continuous acquisition of newly published literature. The resulting *Curated Scientific Literature Repository* provides the validated input for the subsequent *Scientific Knowledge Extraction Module*.

## 4. SCIENTIFIC KNOWLEDGE EXTRACTION MODULE

### 4.1 Framework Overview

The *Scientific Knowledge Extraction Module* constitutes the second component of the proposed *Autonomous Scientific Knowledge Generation Framework*. Building upon the *Curated Scientific Literature Repository* generated in Stage 1 (Section 3), its objective is to transform validated scientific publications into structured, machine-interpretable scientific knowledge. Rather than simply converting PDF documents into plain text, the framework reconstructs scientific observations while preserving the contextual relationships that define their physical meaning. Scientific knowledge is distributed across narrative text, tables, figures, equations, captions, supplementary materials, and methodological discussions. Consequently, effective knowledge extraction requires more than document parsing. It requires identifying scientific entities, recovering their relationships, and preserving the experimental, computational, and physical context associated with each observation. As illustrated in **Figure 3**, the framework combines scientific document processing, hybrid AI-based knowledge extraction, knowledge fusion, semantic mapping, ontology-guided validation, and quality control within a unified workflow. Rule-based algorithms, natural language processing, and large language models operate collaboratively to extract materials, functional properties, numerical values, experimental conditions, computational parameters, and semantic relationships from heterogeneous document content. The extracted information is subsequently fused and validated before being converted into standardized scientific records. A distinguishing feature of the proposed framework is that knowledge extraction is treated as a scientific reasoning problem rather than a document processing task. Individual property values are never interpreted in isolation but remain explicitly associated with their corresponding materials, operating conditions, methodologies, provenance, and confidence estimates. The resulting *Structured Scientific Knowledge Repository* provides the direct input to the *Scientific Data Fusion and Harmonization Module*, where heterogeneous observations are transformed into canonical scientific knowledge suitable for AI-driven reasoning.

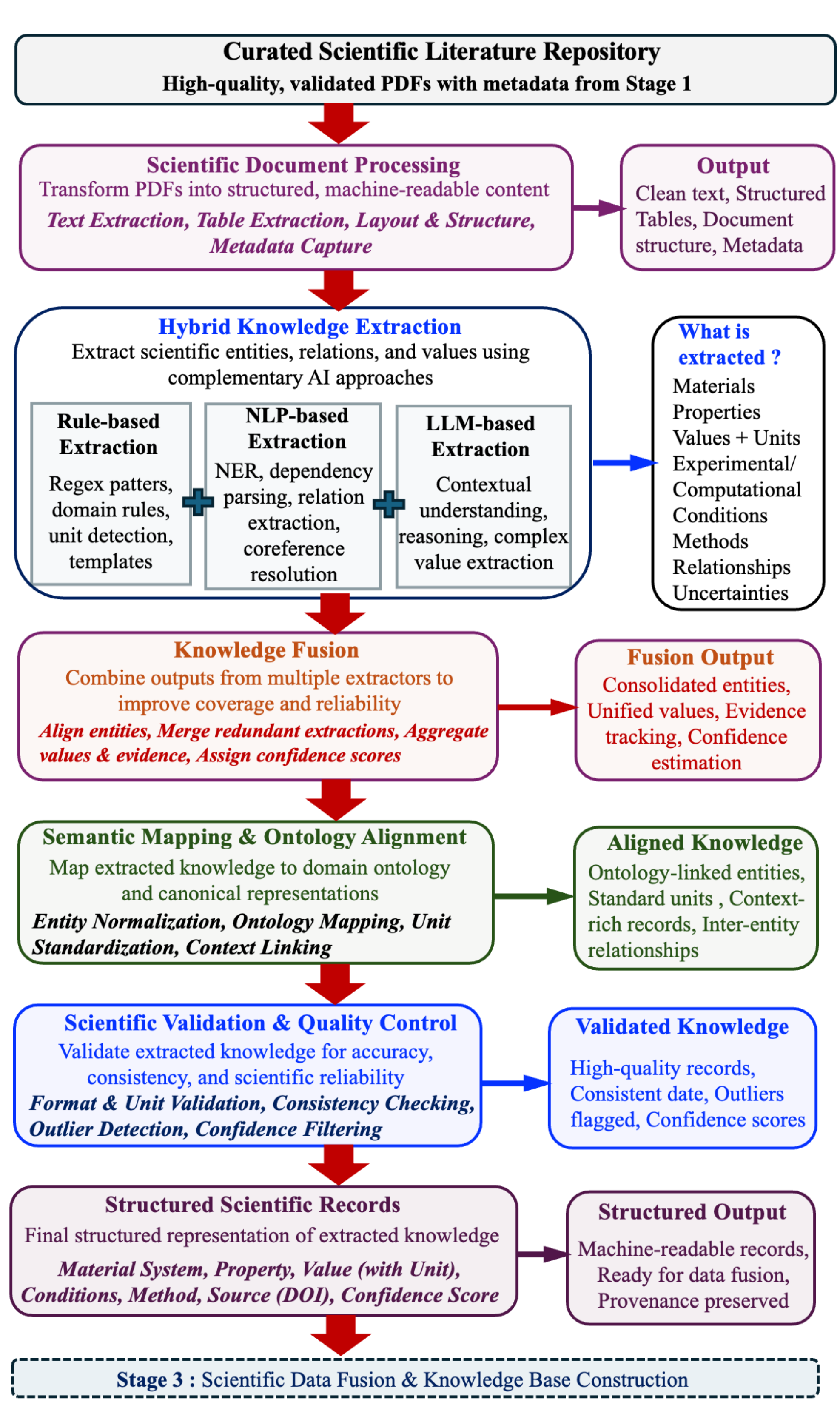

**Figure 3: Scientific Knowledge Extraction Module**

## 4.2 Scientific Document Processing

The first step of the *Scientific Knowledge Extraction Module* converts validated PDF documents into a structured document representation suitable for downstream scientific reasoning. Rather than treating publications as plain text, the framework preserves their logical organization, including narrative text, tables, metadata, document layout, section hierarchy, captions, equations, and supplementary information. Each publication undergoes automated document processing in which full-text content is extracted while maintaining paragraph structure, section boundaries, reading order, and bibliographic metadata. Scientific tables are detected and preserved as structured objects, allowing quantitative measurements to remain associated with their corresponding variables, units, and contextual descriptors. Likewise, document layout and organizational elements provide important contextual cues for distinguishing experimental procedures, computational methods, results, and scientific discussion.

The current implementation focuses primarily on text, tables, metadata, and document structure. However, the modular architecture readily accommodates additional modalities, including scientific figures, plots, microscopy images, schematic diagrams, and multimodal content as computer vision and vision-language foundation models continue to advance. The output of this stage is a structured scientific document representation that preserves both document organization and contextual information, providing the foundation for the hybrid knowledge extraction methods described in the following subsection.

## 4.3 Hybrid Scientific Knowledge Extraction

Scientific knowledge is expressed in diverse forms, ranging from structured numerical values to narrative descriptions, implicit relationships, domain-specific terminology, and contextual scientific reasoning. Consequently, no single extraction strategy is sufficient to recover the full spectrum of knowledge contained within modern scientific publications. The proposed framework therefore employs a hybrid AI-based knowledge extraction architecture that combines complementary reasoning strategies within a unified workflow. As illustrated in **Figure 3**, the framework integrates three complementary paradigms: *rule-based extraction*, *natural language processing*, and *large language model-based reasoning*. Rather than operating independently, these components cooperate to recover structured scientific knowledge from heterogeneous document content while leveraging the strengths of each methodology.

*Rule-based extraction* provides high-precision identification of structured scientific expressions, including numerical values, physical units, chemical compositions, tensor notation, material identifiers, and experimental parameters. NLP complements these capabilities by recognizing scientific entities and semantic relationships embedded in narrative text, while preserving associations among materials, properties, methods, and operating conditions. LLMs provide a higher level of contextual reasoning by integrating information distributed across multiple sections of a publication, resolving implicit relationships, and reconstructing scientific observations that cannot be recovered through deterministic rules or conventional NLP alone. The principal strength of the framework lies in the complementary interaction among these three reasoning paradigms. *Rule-based methods ensure precision*, *NLP captures linguistic structure*, and *LLMs contribute contextual scientific understanding*. Their integration produces a substantially more robust extraction system than any individual methodology operating in isolation.

The objective of the module is therefore not simply to identify entities or numerical values, but to reconstruct structured scientific knowledge in which quantitative measurements remain explicitly associated with materials, experimental conditions, computational methodologies, provenance, and scientific context. These structured knowledge objects provide the foundation for the subsequent stages of knowledge fusion, semantic harmonization, and scientific validation.

### 4.4 Scientific Knowledge Fusion

Scientific knowledge extracted from complex research publications is inherently heterogeneous, and no individual extraction strategy can recover every aspect of the underlying scientific information. Rule-based methods provide highly reliable extraction of structured numerical values, NLP models identify scientific entities and semantic relationships, and LLMs contribute contextual scientific reasoning. Each therefore produces complementary evidence rather than a complete representation of the scientific knowledge contained within a publication. To integrate these complementary outputs, the proposed framework employs a *Scientific Knowledge Fusion Module* that treats each extraction strategy as an independent source of scientific evidence. Rather than selecting a single "best" extractor, the framework systematically compares, aligns, and reconciles the outputs of rule-based algorithms, NLP models, and LLM reasoning to construct a unified scientific representation.

Knowledge fusion operates at multiple semantic levels. Equivalent materials, properties, experimental conditions, computational methods, and physical variables identified by different extraction engines are associated through ontology-guided matching. Numerical values and contextual information are reconciled while preserving units, provenance, confidence estimates, and supporting evidence. Redundant observations are consolidated, complementary information is merged, and potentially conflicting evidence is retained for subsequent scientific validation rather than discarded. A distinguishing feature of the framework is that knowledge fusion preserves scientific evidence throughout the reconciliation process. Every scientific record remains linked to its originating publication, extraction methodology, and confidence estimate, providing complete traceability and transparency for downstream quality assessment. Consequently, the *Scientific Knowledge Fusion* module performs considerably more than data aggregation. By integrating complementary evidence generated through multiple AI reasoning paradigms, it reconstructs coherent, context-aware scientific knowledge that is more complete, reliable, and trustworthy than knowledge produced by any individual extraction method alone. The resulting scientific records provide the foundation for semantic harmonization and construction of the *Unified AI-Ready Scientific Knowledge Base*.

### 4.5 Semantic Mapping and Scientific Validation

Following hybrid extraction and knowledge fusion, the reconstructed scientific observations undergo semantic mapping and scientific validation before being incorporated into the *Structured Scientific Knowledge Repository*. Although complementary extraction engines successfully recover materials, properties, numerical values, and experimental conditions, the resulting records frequently differ in terminology, notation, units, reporting conventions, and contextual representation. Semantic standardization is therefore required before the extracted knowledge can support downstream AI-driven reasoning. The framework addresses this challenge using a domain ontology that defines canonical representations for materials, functional properties, experimental variables, computational methods, and associated scientific relationships. Alternative terminology, abbreviations, tensor notation, units, and reporting conventions are mapped to standardized semantic concepts, enabling heterogeneous information originating from different publications to be interpreted using a consistent scientific vocabulary.

Beyond semantic standardization, the framework reconstructs complete scientific observations by explicitly linking each property value to its corresponding material system, experimental or computational methodology, operating conditions, and provenance. This context-aware representation preserves the physical meaning of every extracted observation rather than treating numerical values as isolated entities. The reconstructed records are subsequently validated for semantic consistency, structural completeness, unit compatibility, and logical coherence. Potential inconsistencies, missing information, or ambiguous extractions are retained together with provenance information, and confidence estimates rather than being discarded, enabling transparent downstream quality assessment. The output of this stage is a *Structured Scientific Knowledge Repository* containing standardized, validated, and context-aware scientific records. These records constitute the direct input to the *Scientific Data Fusion and Harmonization Module*, where heterogeneous observations from multiple publications are reconciled into canonical scientific knowledge.

### 4.6 Batch Scientific Knowledge Generation

The *Scientific Knowledge Extraction Module* is designed for large-scale autonomous processing of scientific literature. Rather than analyzing individual publications in isolation, the framework applies the complete extraction workflow to every document contained within the *Curated Scientific Literature Repository*, enabling efficient knowledge generation from thousands - or ultimately millions - of scientific publications. During batch execution, each publication independently undergoes document processing, hybrid knowledge extraction, knowledge fusion, semantic mapping, and scientific validation. Because individual documents are processed independently, the workflow is inherently parallelizable and readily supports distributed execution and incremental processing as new publications become available.

The validated records generated from all publications are aggregated into a *Structured Scientific Knowledge Repository*. In the current implementation, these records are exported in a standardized tabular format to facilitate interoperability with existing data analysis and machine learning workflows. However, the tabular format is simply an implementation detail. The primary output is a structured, context-aware scientific knowledge repository in which every observation remains linked to its material system, functional property, operating conditions, provenance, and confidence estimate. This repository provides the direct input to the *Scientific Data Fusion and Harmonization Module*, where heterogeneous observations from multiple publications are reconciled into the *Unified AI-Ready Scientific Knowledge Base*.

## 5. SCIENTIFIC DATA FUSION AND HARMONIZATION MODULE

### 5.1 Framework Overview

The *Scientific Data Fusion and Harmonization Module* constitute the third component of the proposed *Autonomous Scientific Knowledge Generation Framework*. Building upon the *Structured Scientific Knowledge Repository* generated in Stage 2 (Section 4), its objective is to transform heterogeneous scientific observations into canonical scientific knowledge that can be consistently interpreted, integrated, and utilized by downstream AI models. Whereas the previous stage reconstructs scientific observations from individual publications, the present stage reconciles those observations across multiple publications, data sources, and reporting conventions. Scientific knowledge is inherently heterogeneous. Identical materials may be reported using different nomenclature, abbreviations, chemical formulas, or common names, while functional properties often appear in different tensor conventions, units, symbols, or measurement protocols. Experimental conditions may likewise be reported numerically, qualitatively, or implicitly. Directly utilizing these heterogeneous observations would introduce redundancy, ambiguity, and inconsistency that limit the reliability of downstream machine learning and generative AI.

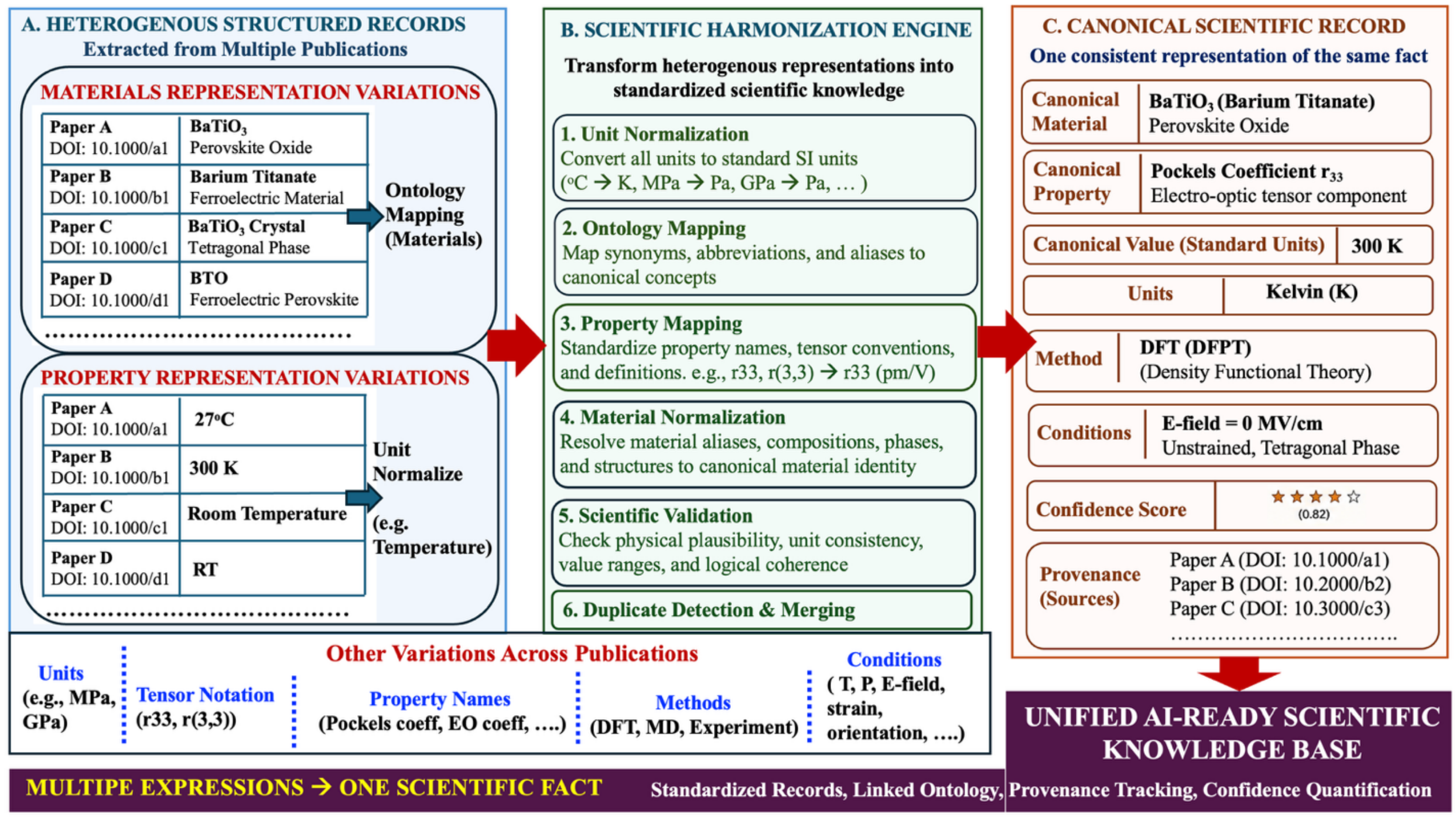

**Figure 4: Scientific Knowledge Harmonization: Multiple Publications → One Canonical Scientific Record**

As illustrated in **Figure 4**, the framework addresses this challenge through ontology-guided semantic harmonization. Structured scientific records undergo semantic mapping, unit normalization, property harmonization, material canonicalization, duplicate resolution, and scientific validation before being incorporated into a unified knowledge infrastructure. Rather than treating these operations as independent data-cleaning tasks, the framework reconstructs canonical scientific knowledge while preserving the experimental context, provenance, and confidence associated with every observation. The output of this stage is a collection of *Canonical Scientific Records* that represent standardized, semantically consistent, and context-aware scientific knowledge. These records constitute the fundamental building blocks of the *Unified AI-Ready Scientific Knowledge Base*, enabling seamless integration with literature, computational simulations, experimental measurements, existing scientific repositories, and future observations.

## 5.2 Scientific Unit Normalization

Scientific measurements are frequently reported using different units, scales, symbols, and reporting conventions, even when describing the same physical quantity. Such heterogeneity complicates the integration of observations from multiple publications and can introduce ambiguity into downstream machine learning and scientific reasoning. The proposed framework addresses this challenge through *Scientific Unit Normalization*, which transforms heterogeneous measurements into canonical physical representations while preserving their scientific context and provenance. Rather than performing simple numerical conversion, the framework interprets equivalent physical quantities using ontology-guided semantic reasoning and maps them to standardized SI-based representations.

As illustrated in **Figure 4**, temperature provides a representative example. *Equivalent operating conditions may be reported as 300 K, 27 °C, room temperature, or ambient conditions*. The framework recognizes these representations as referring to the same physical concept whenever appropriate and converts them into a consistent canonical representation. Similar normalization is performed for functional coefficients, electric fields, pressure, stress, energy, wavelength, frequency, and other physical quantities commonly reported using different units or conventions. Throughout the normalization process, every canonical quantity remains explicitly linked to its original reported value, units, source publication, operating conditions, and confidence estimate. This preserves complete traceability while enabling downstream AI models to learn from semantically consistent scientific knowledge rather than heterogeneous numerical representations. The output of this stage is a collection of canonically represented physical quantities that provides the quantitative foundation for subsequent property harmonization, material semantic normalization, and construction of the *Unified AI-Ready Scientific Knowledge Base*.

### 5.3 Scientific Property Harmonization

While unit normalization establishes consistent numerical representations, AI-driven scientific discovery also requires harmonization of the physical properties themselves. Equivalent properties are often reported using different nomenclature, tensor conventions, abbreviations, symbols, or discipline-specific terminology. Even when numerical values are directly comparable, inconsistent property representations prevent AI systems from recognizing that multiple publications describe the same underlying physical phenomenon. The proposed framework addresses this challenge through *Scientific Property Harmonization*, in which heterogeneous property representations are mapped to canonical scientific definitions using domain ontologies. Rather than treating extracted property names as independent text labels, the framework interprets them as physical concepts that may possess multiple equivalent scientific representations. This ontology-guided interpretation enables observations reported under different conventions to be integrated into a unified semantic representation while preserving their original scientific meaning.

Electro-optic materials provide a representative example[21]. *Across the literature, the same physical property may be described as an electro-optic coefficient, Pockels coefficient, or Pockels constant, or reported using tensor components such as*[21] $r_{33}$, $r_{13}$, $r_{42}$, *and* $r_{51}$. Different publications may further employ alternative tensor conventions or crystallographic reference frames. The framework reconciles these heterogeneous representations into canonical property definitions while preserving the original notation, provenance, and reporting conventions. The resulting *Canonical Scientific Properties* provide standardized semantic representations that remain explicitly linked to their original terminology, source publication, operating conditions, and provenance. Together with normalized physical quantities, these canonical properties establish the semantic foundation for material harmonization, scientific validation, and construction of the *Unified AI-Ready Scientific Knowledge Base*.

### 5.4 Material Semantic Harmonization

A unified scientific knowledge base requires recognizing when different material representations correspond to the same underlying scientific entity. Across the literature, identical materials are frequently reported using chemical formulas, common names, abbreviations, compositional descriptions, crystallographic designations, or application-specific terminology. Without semantic harmonization, these heterogeneous representations would be interpreted as distinct materials, fragmenting the resulting knowledge base and reducing the effectiveness of downstream AI models. The proposed framework addresses this challenge through *Material Semantic Harmonization*, in which extracted material representations are mapped to canonical material identities using domain ontologies and semantic reasoning. Rather than relying solely on lexical similarity, the framework interprets material names together with their compositional, crystallographic, and contextual information to distinguish genuinely equivalent materials from chemically or structurally distinct systems.

As illustrated in **Figure 4**, *the same material may be reported as* $BaTiO_3$*, Barium Titanate,* $BaTiO_3$ *crystal, or BTO*. The framework reconciles these heterogeneous representations into a single canonical material identity while preserving the original aliases, terminology, and source provenance. Similar harmonization is performed for crystallographic phases, compositional variants, doped materials, and other domain-specific material nomenclature. The resulting *Canonical Material Entities* provide a consistent semantic representation of materials while maintaining complete traceability to the original literature. Together with normalized physical quantities and harmonized scientific properties, these canonical material identities establish the semantic foundation of the *Unified AI-Ready Scientific Knowledge Base*.

### 5.5 Scientific Knowledge Fusion and Validation

Following harmonization of physical quantities, scientific properties, and material identities, the framework performs *Scientific Knowledge Fusion and Validation*, where heterogeneous observations extracted from multiple publications are reconciled into coherent and scientifically consistent knowledge. This stage represents the culmination of the harmonization process, transforming independent observations into *canonical scientific records* while preserving their underlying scientific evidence. Scientific literature frequently reports the same physical phenomenon using different terminology, measurement conventions, numerical precision, or experimental methodologies. Consequently, observations that initially appear distinct may describe the same scientific fact, whereas superficially similar observations may correspond to different materials or operating conditions. The proposed framework employs ontology-guided semantic reasoning to distinguish these cases, identifying equivalent observations while preserving the contextual information required to maintain scientific accuracy.

Knowledge fusion combines observations sharing harmonized material identities, canonical property definitions, normalized physical quantities, experimental conditions, and provenance information. Equivalent observations are consolidated into unified canonical records, whereas genuinely distinct observations remain independently represented. Rather than eliminating conflicting evidence, the framework preserves provenance and confidence estimates, enabling downstream AI models to distinguish between experimental variability and genuine scientific disagreement. Each canonical record subsequently undergoes scientific validation to ensure semantic consistency, structural completeness, unit compatibility, and logical coherence. Missing information, ambiguous property assignments, and inconsistent reporting conventions are explicitly retained together with confidence estimates rather than discarded, providing transparent evidence for downstream quality assessment. The output of this stage is a collection of *Canonical Scientific Records*, each representing a validated, semantically consistent, and context-aware scientific observation linked to its original publications, supporting evidence, provenance, and confidence estimate. These canonical records form the fundamental building blocks of the *Unified AI-Ready Scientific Knowledge Base*, providing reliable knowledge for predictive machine learning, generative AI, and autonomous scientific discovery.

### 5.6 Unified AI-Ready Scientific Knowledge Base Generation

The final stage of the *Scientific Data Fusion and Harmonization Module* integrates the validated *Canonical Scientific Records* into a *Unified AI-Ready Scientific Knowledge Base*. Rather than representing a collection of normalized data, the knowledge base organizes scientific observations into interoperable, semantically consistent knowledge objects suitable for AI-driven reasoning. Each canonical record preserves explicit relationships among material identity, functional property, normalized quantitative value, physical units, experimental or computational conditions, provenance, and confidence estimates. Consequently, the knowledge base retains not only numerical measurements but also the scientific context required for reliable interpretation, reproducibility, and downstream machine learning.

In the current implementation, the knowledge base is exported in a standardized tabular format to facilitate interoperability with existing data analysis and machine learning frameworks. However, the tabular representation is merely an implementation format. Conceptually, the framework produces a continuously evolving scientific knowledge infrastructure capable of integrating literature-derived knowledge with computational simulations, experimental measurements, existing scientific repositories, and future observations. The *Unified AI-Ready Scientific Knowledge Base* represents the principal output of the first three stages of the proposed framework and establishes the bridge between autonomous scientific knowledge generation and AI-driven scientific discovery. As illustrated in **Figure 1**, it provides the foundation for predictive scientific intelligence, generative AI, computational validation, and ultimately closed-loop autonomous scientific discovery.

## 6. PROOF-OF-CONCEPT END-TO-END DEMONSTRATION

### 6.1 Demonstration Overview

The objective of this proof-of-concept study is to validate the end-to-end functionality of the proposed *Autonomous Scientific Knowledge Generation Framework* rather than to construct a comprehensive electro-optic materials database. The framework is designed as a scalable architecture capable of continuously processing large and evolving scientific literature collections. Accordingly, the present demonstration focuses on verifying the complete computational workflow - from autonomous literature acquisition to scientific knowledge extraction, semantic harmonization, and AI-ready knowledge-base generation - using a representative subset of electro-optic materials publications. As illustrated in **Figure**

**5**, the *Literature Acquisition Module* autonomously retrieved and validated approximately 1,000 'best' electro-optic publications from multiple scholarly repositories. To evaluate the downstream components of the framework, a representative subset of eight publications was selected for detailed end-to-end analysis. Please check the link under 'code availability' for all files. This subset was chosen solely to demonstrate the complete knowledge generation workflow while preserving full traceability through each intermediate transformation stage.

The proof-of-concept therefore evaluates the framework as an integrated scientific knowledge generation system rather than as an isolated information extraction algorithm. **Figure 5** summarizes the complete end-to-end workflow, whereas **Figure 6** provides representative examples of how individual scientific observations are progressively transformed from original publications into structured scientific records and subsequently into canonical scientific knowledge. Together, these demonstrations validate the framework's ability to autonomously convert heterogeneous scientific literature into a reusable *Unified AI-Ready Scientific Knowledge Base*, establishing the foundation for predictive scientific intelligence, generative AI, and future autonomous scientific discovery.

### 6.2 Validation of Autonomous Literature Acquisition

The first stage of the proposed framework was validated by demonstrating its ability to autonomously acquire and curate domain-specific scientific literature. Beginning with a scientific query on electro-optic materials, the *Literature Acquisition Module* automatically generated ontology-guided search queries, retrieved publications from multiple scholarly repositories, ranked candidate documents according to scientific relevance, acquired available full-text articles, validated metadata, and constructed a curated literature repository. As illustrated in the **Stage 1 of Figure 5**, the automated workflow retrieved and validated approximately 1,000 publications related to electro-optic materials from multiple scientific repositories, including peer-reviewed journals, preprint servers, and other scholarly sources. Throughout the acquisition process, publications underwent automated filtering, relevance assessment, metadata validation, duplicate removal, and quality screening, resulting in a *Curated Scientific Literature Repository* suitable for downstream knowledge extraction.

To validate the subsequent stages of the framework, a *representative subset of eight publications* was selected from the curated repository for detailed end-to-end analysis. It is important to emphasize that this subset was not used to evaluate the performance or scalability of the *Literature Acquisition Module*. Instead, it served exclusively as a representative demonstration corpus for validating the *Scientific Knowledge Extraction* and *Scientific Data Fusion and Harmonization* modules. The complete literature repository remains available for large-scale autonomous processing, whereas the selected subset enables transparent verification of every intermediate transformation stage presented in **Figures 5 and 6**. The successful autonomous acquisition of approximately 1,000 validated publications demonstrates the scalability of the *Literature Acquisition Module* while establishing the foundation for the remainder of the framework. More importantly, it confirms that the proposed architecture can continuously construct and maintain high-quality, domain-specific literature repositories without manual document curation. As illustrated in the lower portion of **Figure 5**, the same acquisition strategy can be directly extended from thousands to hundreds of thousands, and ultimately millions, of scientific publications, providing the knowledge source required for large-scale autonomous scientific knowledge generation.

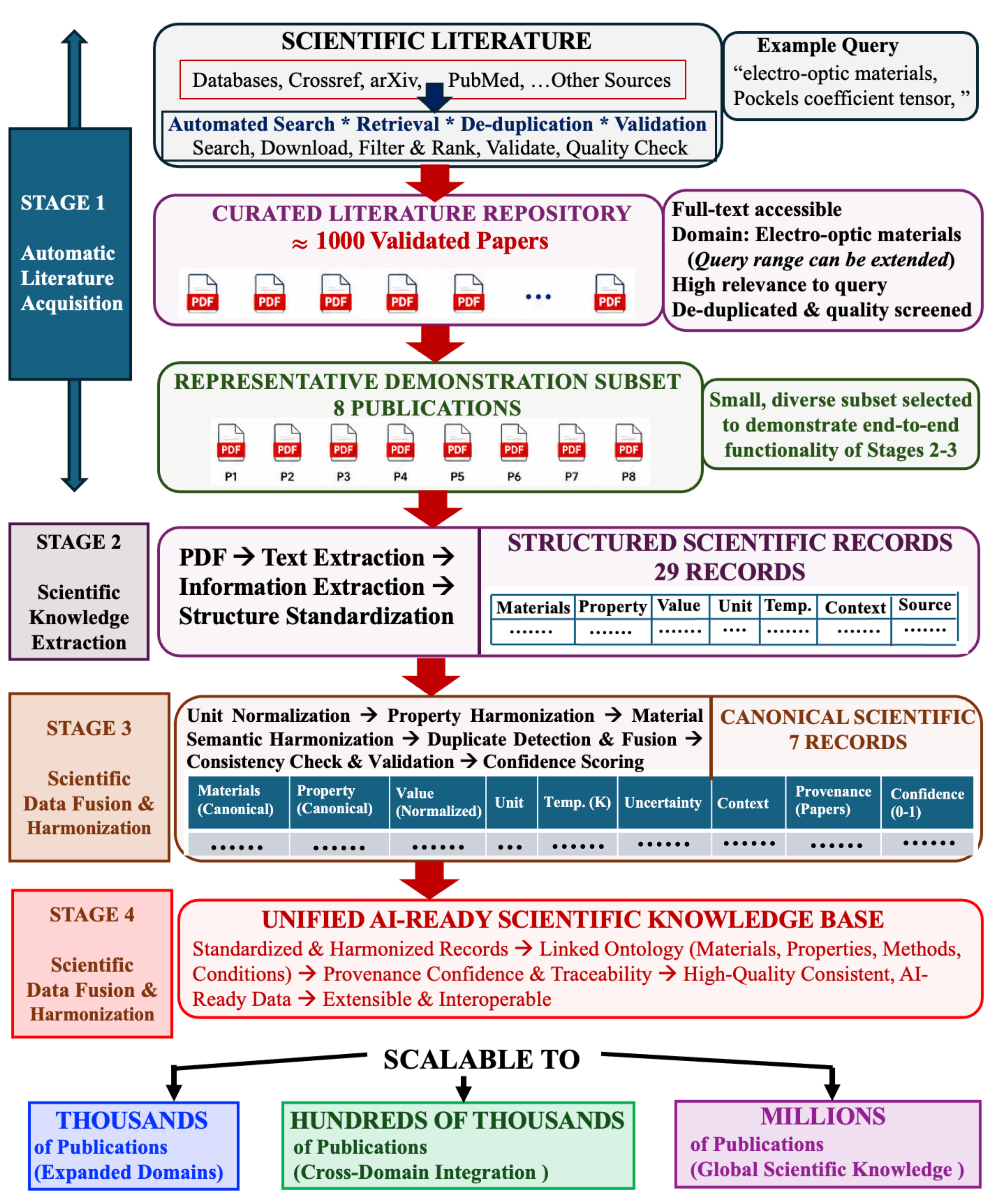

**Figure 5: Proof-Of-Concept Demonstration Using Electro-Optic Materials**

### 6.3 Validation of Scientific Knowledge Extraction

The second stage of the proposed framework was validated by demonstrating its ability to autonomously transform validated scientific publications into structured scientific knowledge. Using the representative demonstration subset introduced in Section 6.2, the *Scientific Knowledge Extraction Module* processed eight electro-optic materials publications through the hybrid extraction workflow described in Section 4. As summarized in **Figure 5**, these publications collectively generated 29 structured scientific records, each representing a context-aware scientific observation suitable for downstream semantic harmonization. Unlike conventional information extraction systems that primarily identify isolated entities or numerical values, the proposed framework reconstructs complete scientific observations by explicitly associating every quantitative measurement with its corresponding material, functional property, operating conditions, provenance, and confidence estimate. Consequently, the extracted records constitute structured scientific knowledge rather than collections of independent textual entities.

The transformation performed during Stage 2 is illustrated in detail in **Figure 6**, which follows representative scientific observations from the original publications to their corresponding structured scientific records. Each example highlights a different aspect of the knowledge extraction process.

**Example 1** demonstrates successful extraction of an electro-optic property from *Venkat_2025_AdvMat.pdf*[21]. The framework correctly identifies the material, tensor component, property type, numerical value, uncertainty, and physical units while preserving the associated provenance. Although the operating temperature is initially unavailable, the structured representation explicitly records this missing information, allowing subsequent harmonization to recover the experimental context.

**Example 2** illustrates a more challenging extraction scenario from *KNO_Zgonik.pdf*[42]. The framework successfully extracts the reported tensor component, numerical value, uncertainty, and operating temperature while preserving the original scientific representation. This example also demonstrates that intermediate structured records intentionally retain the extracted information prior to semantic reconciliation, enabling downstream harmonization to identify inconsistencies and refine the canonical representation without losing traceability to the original publication.

**Example 3** demonstrates extraction of an elastic stiffness coefficient from *BTO_Zgonik.pdf*[43]. In this case, the framework accurately reconstructs the scientific observation while preserving the associated material identity, tensor notation, numerical value, uncertainty, units, and operating conditions. Because the extracted information is already semantically consistent, only minimal modification is required during the subsequent harmonization stage.

Collectively, these examples demonstrate that the *Scientific Knowledge Extraction Module* can reconstruct structured scientific observations across multiple property classes, including electro-optic, piezoelectric, and elastic tensor properties. More importantly, the framework preserves both the quantitative measurements and their associated scientific context, providing the essential information required for downstream semantic harmonization. The generation of 29 structured scientific records from the representative demonstration corpus therefore validates the ability of the proposed framework to autonomously transform heterogeneous scientific literature into structured, machine-interpretable scientific knowledge. These structured records constitute the direct input to Stage 3, where multiple heterogeneous observations are reconciled into canonical scientific knowledge suitable for construction of the *Unified AI-Ready Scientific Knowledge Base*.

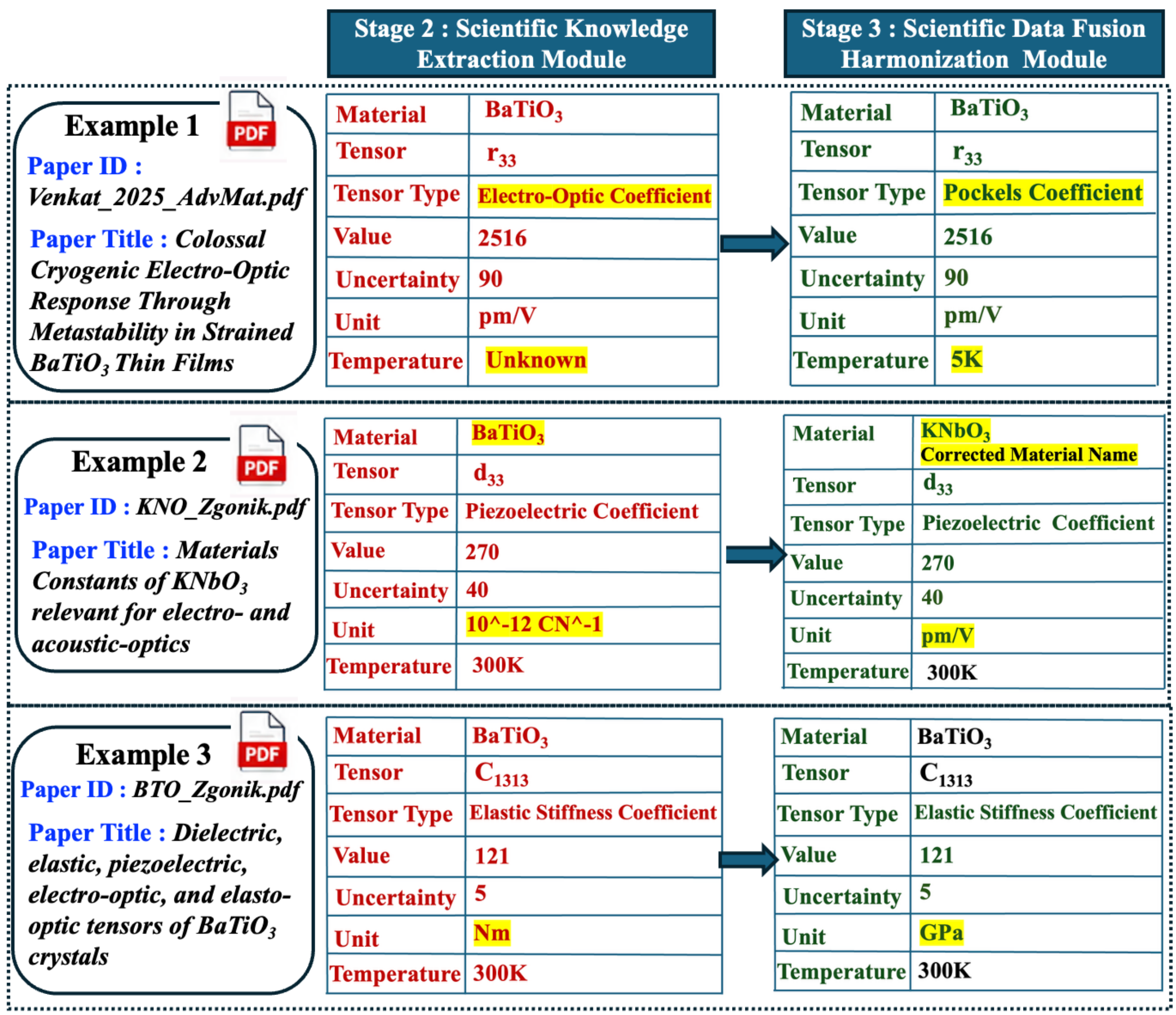

**Figure 6: Proof Of Concept Case Study: Examples of three PDFs**

## 6.4 Validation of Scientific Data Fusion and Harmonization

The third stage of the framework was validated by demonstrating its ability to transform heterogeneous structured scientific records into canonical scientific knowledge. Using the 29 structured scientific records generated during Stage 2, the *Scientific Data Fusion and Harmonization Module* performed ontology-guided unit normalization, property harmonization, material semantic mapping, duplicate identification, scientific validation, and confidence assessment. As summarized in **Figure 5**, this process produced 7 *Canonical Scientific Records*, representing the consolidated scientific knowledge contained within the representative demonstration dataset. Importantly, the reduction from 29 structured records to 7 canonical records does not represent information loss or data filtering. Rather, it reflects the successful reconciliation of multiple heterogeneous descriptions of the same scientific observations. Scientific publications frequently report identical materials, properties, operating conditions, or measurement conventions using different terminology, abbreviations, units, tensor notation, or reporting styles. The harmonization module identifies these semantically equivalent observations and integrates them into unified canonical representations while preserving complete provenance and traceability. The transformation performed during Stage 3 is illustrated in the **right-hand column of Figure 6**, which demonstrates how structured scientific records are refined into canonical scientific records.

**Example 1** illustrates ***context completion and semantic refinement***. The structured record successfully identifies the electro-optic tensor component, numerical value, uncertainty, and physical units from *Venkat_2025_AdvMat.pdf*[21]. During harmonization, the property description is standardized from *Electro-Optic Coefficient* to the canonical term *Pockels Coefficient*, while the missing operating temperature is supplemented using contextual information recovered during semantic reasoning. The resulting canonical record therefore contains a more complete scientific representation while preserving the original provenance.

**Example 2** demonstrates ***semantic correction and unit harmonization***. Although the structured record contains the extracted tensor component, numerical value, uncertainty, and operating temperature, the material identity initially remains inconsistent with the source publication. During Stage 3, ontology-guided semantic reasoning correctly maps the material to $KNbO_3$, harmonizes the reported units into a canonical representation, and preserves the original scientific observation together with complete provenance information. This example illustrates how semantic harmonization can identify and correct inconsistencies that cannot be resolved during document-level extraction alone.

**Example 3** illustrates ***scientific validation of an already consistent record***. The structured record accurately captures the elastic stiffness coefficient, material identity, numerical value, uncertainty, operating temperature, and associated units from *BTO_Zgonik.pdf*. Since the extracted information is already semantically consistent, Stage 3 preserves the observation with only minimal standardization of terminology and units. This example demonstrates that the harmonization framework performs validation rather than unnecessary modification when the extracted scientific knowledge is already internally consistent.

Collectively, these examples illustrate three complementary capabilities of the *Scientific Data Fusion and Harmonization Module*: semantic refinement of extracted knowledge, ontology-guided correction of heterogeneous scientific representations, and validation of already consistent scientific observations. More importantly, they demonstrate that the framework performs considerably more than conventional data cleaning. It reconstructs canonical scientific knowledge by integrating semantic reasoning, ontology-guided harmonization, provenance preservation, and confidence assessment into a unified workflow. The successful generation of *7 Canonical Scientific Records* from *29 heterogeneous structured records* therefore validates the framework's ability to autonomously reconcile distributed scientific evidence into semantically consistent, context-aware, and AI-ready scientific knowledge. These canonical records constitute the fundamental building blocks of the *Unified AI-Ready Scientific Knowledge Base*, providing the reliable knowledge infrastructure required for predictive scientific intelligence, generative AI, and closed-loop autonomous scientific discovery.

## 7. TOWARD AUTONOMOUS SCIENTIFIC DISCOVERY

### 7.1 Predictive Scientific Intelligence

The *Unified AI-Ready Scientific Knowledge Base* established by the proposed framework provides the foundation for predictive scientific intelligence, enabling AI systems to learn quantitative relationships among materials composition, structure, processing history, operating conditions, and functional performance directly from harmonized scientific knowledge. Unlike conventional machine learning workflows that rely on isolated datasets collected for individual studies, the proposed framework continuously integrates literature-derived knowledge with computational simulations, experimental measurements, and future scientific observations, providing a richer and more comprehensive representation of materials behavior. This integrated knowledge enables predictive AI models to rapidly estimate functional properties of previously unexplored materials while accounting for the experimental and computational context in which those properties were obtained. Because each canonical scientific record preserves material identity, operating conditions, provenance, uncertainty, and semantic relationships, predictive models can learn from scientifically consistent observations rather than fragmented numerical datasets. Such context-aware learning improves both prediction reliability and interpretability while facilitating knowledge transfer across diverse materials systems.

The resulting predictive capability extends beyond conventional property regression. AI models trained on the generated knowledge base can rapidly screen multidimensional design spaces, identify relationships among composition, processing conditions, and performance, quantify prediction uncertainty, and prioritize promising candidate materials for further computational or experimental investigation. Consequently, predictive scientific intelligence transforms the generated knowledge base into an effective decision-support system, enabling researchers to focus computational and experimental resources on the most promising regions of the design space. An important characteristic of the proposed framework is that predictive intelligence continuously improves as the knowledge base evolves. Newly published literature, computational simulations, laboratory experiments, and other trusted scientific observations are automatically incorporated through the autonomous knowledge generation workflow, enabling predictive models to be periodically retrained using increasingly comprehensive scientific knowledge. In this manner, the framework establishes a continuously learning predictive ecosystem in which expanding scientific knowledge directly translates into progressively more accurate and reliable scientific predictions.

Although the present work focuses on autonomous scientific knowledge generation rather than predictive model development, the generated *Unified AI-Ready Scientific Knowledge Base* provides the essential information infrastructure required for future AI-driven property prediction and scientific decision support. It therefore represents the first step toward autonomous scientific discovery, where continuously evolving scientific knowledge enables increasingly intelligent computational reasoning and accelerates the identification of promising materials for subsequent validation.

### 7.2 Generative Scientific Intelligence

While predictive scientific intelligence enables AI systems to estimate the properties of existing materials, the next step toward autonomous scientific discovery is the ability to generate entirely new scientific hypotheses. The *Unified AI-Ready Scientific Knowledge Base* provides the foundation for this capability by enabling *Generative Scientific Intelligence*, in which AI systems perform inverse scientific reasoning to identify candidate materials, processing strategies, and operating conditions that satisfy predefined functional objectives. Unlike conventional generative approaches that rely on isolated computational datasets or narrowly defined design spaces, the proposed framework integrates scientific knowledge derived from literature, computational simulations, experimental measurements, and existing scientific repositories into a unified semantic representation. This comprehensive knowledge enables generative AI to learn not only relationships between composition and functional properties but also the influence of synthesis routes, processing history, operating conditions, experimental methodologies, and scientific constraints. Consequently, the search space extends beyond materials composition to encompass the broader scientific context that governs materials performance.

Within this framework, inverse scientific design begins by specifying a desired target - for example, a high electro-optic coefficient, enhanced dielectric response, improved mechanical stability, or application-specific operating conditions. Rather than searching existing databases for similar materials, the AI system reasons across the accumulated scientific knowledge to propose candidate materials together with associated processing conditions, operating environments, or external control parameters that are predicted to satisfy the target objective. In this manner, the knowledge base serves not only as a repository of existing observations but also as a platform for generating new scientific hypotheses. An important advantage of this approach is that every generated hypothesis remains grounded in harmonized scientific knowledge. Because candidate designs are derived from canonical scientific records that preserve provenance, uncertainty, experimental context, and semantic relationships, the resulting predictions remain scientifically interpretable and traceable rather than functioning as isolated AI-generated suggestions. This transparency facilitates subsequent computational verification and experimental validation while providing researchers with insight into the scientific evidence supporting each proposed candidate.

Although the present work does not implement large-scale inverse materials design, it establishes the essential knowledge infrastructure required for future generative AI systems. By transforming heterogeneous scientific literature into a continuously evolving *Unified AI-Ready Scientific Knowledge Base*, the proposed framework enables AI to progress from predicting known materials behavior to generating scientifically informed candidate materials and testable hypotheses. This transition represents a critical step toward autonomous scientific discovery, where continuously expanding scientific knowledge drives increasingly intelligent materials design and scientific innovation.

### 7.3 Closed-Loop Autonomous Scientific Discovery

The ultimate objective of the proposed framework extends beyond autonomous literature mining, scientific knowledge extraction, or AI-assisted prediction considered individually. Its long-term vision is the establishment of a closed-loop autonomous scientific discovery ecosystem, in which scientific knowledge is continuously acquired, interpreted, validated, and reused to improve subsequent scientific discovery. Rather than operating as independent computational components, knowledge generation, predictive scientific intelligence, generative scientific intelligence, computational modeling, and experimental validation become interconnected elements of a continuously learning scientific system. The process begins with the *Unified AI-Ready Scientific Knowledge Base*, which serves as the central knowledge infrastructure of the framework. Predictive AI models utilize this knowledge to identify relationships among materials composition, processing conditions, operating environments, and functional performance, providing rapid estimates of materials behavior and guiding scientific decision making. Building upon these predictions, generative AI performs inverse scientific reasoning by proposing candidate materials, processing strategies, and operating conditions that satisfy predefined functional objectives. These AI-generated hypotheses subsequently undergo computational validation through first-principles calculations, molecular simulations, phase-field modeling, finite element analysis, or other appropriate modeling approaches before being evaluated experimentally.

A defining characteristic of the proposed framework is that validation is not the end of the discovery process. Newly generated computational and experimental observations are automatically incorporated into the knowledge base through the same knowledge generation and harmonization workflow described in the preceding sections. Consequently, every new scientific observation - whether obtained from literature, simulation, or experiment - becomes additional training knowledge that continuously expands and refines the underlying scientific understanding. As the knowledge base evolves, predictive models become more accurate, generative models become more reliable, and the overall discovery process progressively improves. This continuous feedback mechanism fundamentally distinguishes the proposed framework from conventional sequential research workflows. Traditional scientific investigations often proceed through isolated cycles of literature review, computation, experimentation, and publication. In contrast, the proposed architecture transforms each completed investigation into new scientific knowledge that immediately becomes available for future AI-driven reasoning. Scientific literature, computational simulations, experimental measurements, and AI-generated hypotheses therefore form a self-improving scientific ecosystem in which every discovery contributes to the next.

Although the present work focuses on establishing the knowledge-generation infrastructure, the proposed framework provides a scalable foundation for future autonomous scientific discovery across diverse scientific disciplines. As the *Unified AI-Ready Scientific Knowledge Base* expands from thousands to millions of scientific observations, increasingly sophisticated predictive and generative AI systems can operate on progressively richer scientific knowledge, enabling more accurate prediction, more efficient inverse design, faster computational screening, and more targeted experimental validation. In this vision, scientific knowledge continuously improves scientific intelligence, scientific intelligence accelerates scientific discovery, and every new discovery enriches the knowledge base that drives the next generation of innovation.

## 8. CONCLUSION AND FUTURE OUTLOOK

This work presents an *Autonomous Scientific Knowledge Generation Framework* that systematically transforms unstructured scientific literature into a *Unified AI-Ready Scientific Knowledge Base*, establishing the information infrastructure required for next-generation AI-driven scientific discovery. Rather than treating literature retrieval, information extraction, semantic harmonization, and knowledge integration as independent tasks, the proposed framework unifies these processes within a single computational architecture that progressively converts distributed scientific observations into structured, semantically consistent, and context-aware scientific knowledge. The proof-of-concept demonstration using electro-optic materials validates the complete end-to-end workflow. Beginning with autonomous acquisition of approximately 1,000 scientific publications, the framework successfully generated structured scientific records from a representative validation subset and subsequently reconciled these heterogeneous observations into canonical scientific knowledge. More importantly, the demonstration illustrates the central concept of the proposed framework: *scientific literature is transformed from a collection of human-readable documents into machine-interpretable scientific knowledge that preserves quantitative measurements together with their associated physical context, provenance, and semantic relationships.*

Although electro-optic materials were selected as a representative benchmark because of their complex tensor properties, multidimensional operating conditions, and diverse reporting conventions, the framework is inherently domain agnostic. By modifying the underlying scientific ontology, retrieval strategy, and extraction schema, the same computational architecture can be readily adapted to batteries, catalysis, semiconductors, quantum materials, polymers, biomaterials, nanomedicine, and numerous other scientific disciplines. Looking forward, the proposed framework provides a scalable foundation for future AI-driven scientific discovery. As the knowledge base expands through autonomous literature acquisition together with computational simulations, high-throughput calculations, laboratory experiments, and future scientific observations, increasingly sophisticated predictive and generative AI models can be trained using richer and continuously evolving scientific knowledge. Future developments will include multimodal scientific knowledge extraction from figures, graphs, microscopy images, spectra, and supplementary materials, as well as seamless integration with high-throughput simulations, laboratory automation, and autonomous experimentation.

More broadly, this work represents a shift in how artificial intelligence can interact with scientific knowledge. Instead of viewing AI solely as a tool for analyzing existing datasets, we envision AI as an active participant throughout the scientific discovery process - from autonomous acquisition and interpretation of scientific knowledge to predictive reasoning, inverse design, computational validation, experimental feedback, and continuous learning. By transforming the world's scientific literature into an evolving AI-ready scientific knowledge infrastructure, the proposed framework establishes a pathway toward a future in which scientific knowledge continuously improves scientific intelligence, and scientific intelligence continuously accelerates scientific discovery.

## RESOURE AVAILABILITY

### Author Contact

Further information and requests for resources and materials should be directed to and will be fulfilled by the author, Dibakar Datta (ddlab@njit.edu)

### Materials availability

No materials were synthesized in this study.

### Data and code availability

- Data can be obtained by request from the author.
- Code and test pdfs are available in Google Drive: https://drive.google.com/drive/folders/1DDPBGw_eTWViJedSPIf20a462bwJYz_z?usp=drive_link
- Any additional information required to reanalyze the data reported in this paper in available from the author upon request.

## AUTHOR CONTRIBUTIONS

D.D. conceived the project, performed all work, and wrote the manuscript.

## DECLARATION OF INTERESTS

The author has no conflict of interest to declare.

## ACKNOWLEDGEMENTS

The work is supported by the National Science Foundation (NSF) DMREF Program (award number # 2522898). The authors acknowledge Advanced Cyberinfrastructure Coordination Ecosystem: Service & Support (ACCESS) for the computational facilities (award number DMR180013 and MAT250009).

## REFERENCES

(1) Wang, H.; Fu, T.; Du, Y.; Gao, W.; Huang, K.; Liu, Z.; Chandak, P.; Liu, S.; Van Katwyk, P.; Deac, A.; et al. Scientific discovery in the age of artificial intelligence. *Nature* **2023**, *620* (7972), 47-60. DOI: 10.1038/s41586-023-06221-2.

(2) Fui-Hoon Nah, F.; Zheng, R.; Cai, J.; Siau, K.; Chen, L. Generative AI and ChatGPT: Applications, challenges, and AI-human collaboration. *Journal of Information Technology Case and Application Research* **2023**, *25* (3), 277-304. DOI: 10.1080/15228053.2023.2233814.

(3) Karniadakis, G. E.; Kevrekidis, I. G.; Lu, L.; Perdikaris, P.; Wang, S.; Yang, L. Physics-informed machine learning. *Nature Reviews Physics* **2021**, *3* (6), 422-440. DOI: 10.1038/s42254-021-00314-5.

(4) Hirschberg, J.; Manning, C. D. Advances in natural language processing. *Science* **2015**, *349* (6245), 261-266. DOI: 10.1126/science.aaa8685.

(5) LeCun, Y.; Bengio, Y.; Hinton, G. Deep learning. *Nature* **2015**, *521* (7553), 436-444. DOI: 10.1038/nature14539.

(6) Zhu, J.; Ren, Y.; Zhou, W.; Xu, J.; Niu, Z.; Zhan, S.; Ma, W. LLM-mambaformer: Integrating mamba and transformer for crystalline solids properties prediction. *Materials Today Communications* **2025**, *44*. DOI: 10.1016/j.mtcomm.2025.112029.

(7) Datta, J. N., A.;  Koratkar, N.; Datta, D. Generative AI for discovering porous oxide materials for next-generation energy storage. *Cell Reports Physical Science* **2025**, *6* (102665). DOI: 10.1016/j.xcrp.2025.102665.

(8) Curtarolo, S.; Hart, G. L.; Nardelli, M. B.; Mingo, N.; Sanvito, S.; Levy, O. The high-throughput highway to computational materials design. *Nat Mater* **2013**, *12* (3), 191-201. DOI: 10.1038/nmat3568.

(9) Matsuzaka, Y.; Yashiro, R. AI-Based Computer Vision Techniques and Expert Systems. *Ai* **2023**, *4* (1), 289-302. DOI: 10.3390/ai4010013.

(10) Koh, H. Y.; Zheng, Y.; Yang, M.; Arora, R.; Webb, G. I.; Pan, S.; Li, L.; Church, G. M. AI-driven protein design. *Nature Reviews Bioengineering* **2025**, *3* (12), 1034-1056. DOI: 10.1038/s44222-025-00349-8.

(11) Thirunavukarasu, A. J.; Ting, D. S. J.; Elangovan, K.; Gutierrez, L.; Tan, T. F.; Ting, D. S. W. Large language models in medicine. *Nat Med* **2023**, *29* (8), 1930-1940. DOI: 10.1038/s41591-023-02448-8.

(12) Zeni, C.; Pinsler, R.; Zugner, D.; Fowler, A.; Horton, M.; Fu, X.; Wang, Z.; Shysheya, A.; Crabbe, J.; Ueda, S.; et al. A generative model for inorganic materials design. *Nature* **2025**, *639* (8055), 624-632. DOI: 10.1038/s41586-025-08628-5.

(13) Noh, J.; Kim, J.; Stein, H. S.; Sanchez-Lengeling, B.; Gregoire, J. M.; Aspuru-Guzik, A.; Jung, Y. Inverse Design of Solid-State Materials via a Continuous Representation. *Matter* **2019**, *1* (5), 1370-1384. DOI: 10.1016/j.matt.2019.08.017.

(14) Liu, Y.; Yang, Z.; Yu, Z.; Liu, Z.; Liu, D.; Lin, H.; Li, M.; Ma, S.; Avdeev, M.; Shi, S. Generative artificial intelligence and its applications in materials science: Current situation and future perspectives. *Journal of Materiomics* **2023**, *9* (4), 798-816. DOI: 10.1016/j.jmat.2023.05.001.

(15) Stach, E.; DeCost, B.; Kusne, A. G.; Hattrick-Simpers, J.; Brown, K. A.; Reyes, K. G.; Schrier, J.; Billinge, S.; Buonassisi, T.; Foster, I.; et al. Autonomous experimentation systems for materials development: A community perspective. *Matter* **2021**, *4* (9), 2702-2726. DOI: 10.1016/j.matt.2021.06.036.

(16) Häse, F.; Roch, L. M.; Aspuru-Guzik, A. Next-Generation Experimentation with Self-Driving Laboratories. *Trends in Chemistry* **2019**, *1* (3), 282-291. DOI: 10.1016/j.trechm.2019.02.007.

(17) Jain, A.; Ong, S. P.; Hautier, G.; Chen, W.; Richards, W. D.; Dacek, S.; Cholia, S.; Gunter, D.; Skinner, D.; Ceder, G.; et al. Commentary: The Materials Project: A materials genome approach to accelerating materials innovation. *APL Materials* **2013**, *1* (1). DOI: 10.1063/1.4812323.

(18) Kirklin, S.; Saal, J. E.; Meredig, B.; Thompson, A.; Doak, J. W.; Aykol, M.; Rühl, S.; Wolverton, C. The Open Quantum Materials Database (OQMD): assessing the accuracy of DFT formation energies. *npj Computational Materials* **2015**, *1* (1). DOI: 10.1038/npjcompumats.2015.10.

(19) Curtarolo, S.; Setyawan, W.; Hart, G. L. W.; Jahnatek, M.; Chepulskii, R. V.; Taylor, R. H.; Wang, S.; Xue, J.; Yang, K.; Levy, O.; et al. AFLOW: An automatic framework for high-throughput materials discovery. *Computational Materials Science* **2012**, *58*, 218-226. DOI: 10.1016/j.commatsci.2012.02.005.

(20) Hellenbrandt, M. The Inorganic Crystal Structure Database (ICSD)—Present and Future. *Crystallography Reviews* **2014**, *10* (1), 17-22. DOI: 10.1080/08893110410001664882.

(21) Suceava, A.; Hazra, S.; Ross, A.; Philippi, I. R.; Sotir, D.; Brower, B.; Ding, L.; Zhu, Y.; Zhang, Z.; Sarkar, H.; et al. Colossal Cryogenic Electro-Optic Response Through Metastability in Strained BaTiO(3) Thin Films. *Adv Mater* **2026**, *38* (3), e07564. DOI: 10.1002/adma.202507564.

(22) Liang, C.; Rouzhahong, Y.; Ye, C.; Li, C.; Wang, B.; Li, H. Material symmetry recognition and property prediction accomplished by crystal capsule representation. *Nat Commun* **2023**, *14* (1), 5198. DOI: 10.1038/s41467-023-40756-2.

(23) Chen, L.-Q. Phase-Field Models for Microstructure Evolution. *Annual Review of Materials Research* **2002**, *32* (1), 113-140. DOI: 10.1146/annurev.matsci.32.112001.132041.

(24) Choi, K. J.; Biegalski, M.; Li, Y. L.; Sharan, A.; Schubert, J.; Uecker, R.; Reiche, P.; Chen, Y. B.; Pan, X. Q.; Gopalan, V.; et al. Enhancement of ferroelectricity in strained BaTiO3 thin films. *Science* **2004**, *306* (5698), 1005-1009. DOI: 10.1126/science.1103218.

(25) Datta, D. Electro-Chemo-Mechanical Modeling of Multiscale Active Materials for Next-Generation Energy Storage: Opportunities and Challenges. *Jom* **2024**, *76* (3), 1110-1130. DOI: 10.1007/s11837-023-06335-y.

(26) Dixit, S.; Kumar, A.; Srinivasan, K.; Vincent, P.; Ramu Krishnan, N. Advancing genome editing with artificial intelligence: opportunities, challenges, and future directions. *Front Bioeng Biotechnol* **2023**, *11*, 1335901. DOI: 10.3389/fbioe.2023.1335901.

(27) Lei, G.; Docherty, R.; Cooper, S. J. Materials science in the era of large language models: a perspective. *Digital Discovery* **2024**, *3* (7), 1257-1272. DOI: 10.1039/d4dd00074a.

(28) Chen, F.; Zhu, Y.; Liu, S.; Qi, Y.; Hwang, H. Y.; Brandt, N. C.; Lu, J.; Quirin, F.; Enquist, H.; Zalden, P.; et al. Ultrafast terahertz-field-driven ionic response in ferroelectric BaTiO3. *Physical Review B* **2016**, *94* (18). DOI: 10.1103/PhysRevB.94.180104.

(29) Cohen, R. E. Origin of ferroelectricity in perovskite oxides. *Nature* **1992**, *358* (6382), 136-138. DOI: 10.1038/358136a0.

(30) Eltes, F.; Villarreal-Garcia, G. E.; Caimi, D.; Siegwart, H.; Gentile, A. A.; Hart, A.; Stark, P.; Marshall, G. D.; Thompson, M. G.; Barreto, J.; et al. An integrated optical modulator operating at cryogenic temperatures. *Nat Mater* **2020**, *19* (11), 1164-1168. DOI: 10.1038/s41563-020-0725-5.

(31) Gopalan, V.; Gupta, M. C. Observation of internal field in LiTaO3 single crystals: Its origin and time-temperature dependence. *Applied Physics Letters* **1996**, *68* (7), 888-890. DOI: 10.1063/1.116220.

(32) Moradifar, P.; Liu, Y.; Shi, J.; Siukola Thurston, M. L.; Utzat, H.; van Driel, T. B.; Lindenberg, A. M.; Dionne, J. A. Accelerating Quantum Materials Development with Advances in Transmission Electron Microscopy. *Chem Rev* **2023**, *123* (23), 12757-12794. DOI: 10.1021/acs.chemrev.2c00917.

(33) Schlom, D. G.; Chen, L. Q.; Pan, X.; Schmehl, A.; Zurbuchen, M. A. A Thin Film Approach to Engineering Functionality into Oxides. *Journal of the American Ceramic Society* **2008**, *91* (8), 2429-2454. DOI: 10.1111/j.1551-2916.2008.02556.x.

(34) Hendricks, G.; Tkaczyk, D.; Lin, J.; Feeney, P. Crossref: The sustainable source of community-owned scholarly metadata. *Quantitative Science Studies* **2020**, *1* (1), 414-427. DOI: 10.1162/qss_a_00022.

(35) Ginsparg, P. ArXiv at 20. *Nature* **2011**, *476* (7359), 145-147. DOI: 10.1038/476145a.

(36) White, J. PubMed 2.0. *Med Ref Serv Q* **2020**, *39* (4), 382-387. DOI: 10.1080/02763869.2020.1826228.

(37) Birkle, C.; Pendlebury, D. A.; Schnell, J.; Adams, J. Web of Science as a data source for research on scientific and scholarly activity. *Quantitative Science Studies* **2020**, *1* (1), 363-376. DOI: 10.1162/qss_a_00018.

(38) Burnham, J. F. Scopus database: a review. *Biomed Digit Libr* **2006**, *3*, 1. DOI: 10.1186/1742-5581-3-1.

(39) Fricke, S. Semantic Scholar. *Journal of the Medical Library Association* **2018**, *106* (1). DOI: 10.5195/jmla.2018.280.

(40) Haddaway, N. R.; Collins, A. M.; Coughlin, D.; Kirk, S. The Role of Google Scholar in Evidence Reviews and Its Applicability to Grey Literature Searching. *PLoS One* **2015**, *10* (9), e0138237. DOI: 10.1371/journal.pone.0138237.

(41) Chandrakar, R. Digital object identifier system: an overview. *The Electronic Library* **2006**, *24* (4), 445-452. DOI: 10.1108/02640470610689151.

(42) Zgonik, M.; Schlesser, R.; Biaggio, I.; Voit, E.; Tscherry, J.; Günter, P. Materials constants of KNbO3 relevant for electro- and acousto-optics. *Journal of Applied Physics* **1993**, *74* (2), 1287-1297. DOI: 10.1063/1.354934.

(43) Zgonik, M.; Bernasconi, P.; Duelli, M.; Schlesser, R.; Gunter, P.; Garrett, M. H.; Rytz, D.; Zhu, Y.; Wu, X. Dielectric, elastic, piezoelectric, electro-optic, and elasto-optic tensors of BaTiO3 crystals. *Phys Rev B Condens Matter* **1994**, *50* (9), 5941-5949. DOI: 10.1103/physrevb.50.5941.